\documentclass[9pt,letterpaper,conference]{IEEEtran}

\usepackage{cite}

\usepackage{amsmath,amssymb,epsfig}
\usepackage{bm}
\usepackage[usenames, dvipsnames]{color}

\usepackage[font=small,labelfont=bf]{caption}
\usepackage[font=small]{subfig}

\makeatletter
\def\footnoterule{
  	\kern-3\p@
	\hrule\@width.4\columnwidth
	\kern2.6\p@}
\makeatother

\IEEEoverridecommandlockouts

\DeclareMathOperator*{\argminA}{arg\,min}

\newcommand{\bx}{\ensuremath{\bold{x}}}
\newcommand{\bbR}{\ensuremath{\mathbb{R}}}
\newcommand{\bbQ}{\ensuremath{\mathbb{Q}}}
\newcommand{\bbC}{\ensuremath{\mathbb{C}}}

\newcommand{\cF}{\ensuremath{\mathcal{F}}}
\newcommand{\bbE}{\ensuremath{\mathbb{E}}}
\newcommand{\bM}{\ensuremath{\bold{M}}}

\newcommand{\bW}{\ensuremath{\bold{W}}}
\newcommand{\bF}{\ensuremath{\bold{F}}}
\newcommand{\bC}{\ensuremath{\bold{C}}}
\newcommand{\bN}{\ensuremath{\bold{N}}}
\newcommand{\bA}{\ensuremath{\bold{A}}}
\newcommand{\bJ}{\ensuremath{\bold{J}}}
\newcommand{\bI}{\ensuremath{\bold{I}}}
\newcommand{\bzero}{\ensuremath{\bold{0}}}

\newcommand{\btheta}{\ensuremath{\boldsymbol{\theta}}}

\newcommand{\bsig}{\ensuremath{\boldsymbol{\sigma}}}
\newcommand{\bq}{\ensuremath{\bold{q}}}
\newcommand{\bh}{\ensuremath{\bold{h}}}
\newcommand{\ba}{\ensuremath{\bold{a}}}
\newcommand{\bb}{\ensuremath{\bold{b}}}
\newcommand{\bmm}{\ensuremath{\bold{m}}}
\newcommand{\bn}{\ensuremath{\bold{n}}}
\newcommand{\bu}{\ensuremath{\bold{u}}}
\newcommand{\bv}{\ensuremath{\bold{v}}}
\newcommand{\br}{\ensuremath{\bold{r}}}
\newcommand{\vph}{\ensuremath{\varphi}}
\newcommand{\us}{\ensuremath{\underline{s}}}
\newcommand{\Gk}{\ensuremath{G^{(k)}}}
\newcommand{\Go}{\ensuremath{G^{(1)}}}

\newcommand{\fik}{\ensuremath{f_i^{(k)}}}
\newcommand{\fikp}{\ensuremath{f_i^{(k+1)}}}
\newcommand{\fio}{\ensuremath{f_i^{(1)}}}

\newcommand{\fnk}{\ensuremath{f_0^{(k)}}}
\newcommand{\fok}{\ensuremath{f_1^{(k)}}}

\newcommand{\fokm}{\ensuremath{f_1^{(k)}}}

\newcommand{\phnk}{\ensuremath{\vph_0^{(k)}}}
\newcommand{\phok}{\ensuremath{\vph_1^{(k)}}}
\newcommand{\munok}{\ensuremath{\mu_{01}^{(k)}}}
\newcommand{\munoo}{\ensuremath{\mu_{01}^{(1)}}}
\newcommand{\deltanok}{\ensuremath{\delta_{01}^{(k)}}}
\newcommand{\snk}{\ensuremath{s_{0}^{(k)}}}

\newcommand{\sok}{\ensuremath{s_{1}^{(k)}}}
\newcommand{\rnok}{\ensuremath{r_{01}^{(k)}}}

\newcommand{\psink}{\ensuremath{\psi_0^{(k)}}}
\newcommand{\psiok}{\ensuremath{\psi_1^{(k)}}}
\newcommand{\epsino}{\ensuremath{e^{-j\psi_0^{(1)}}}}
\newcommand{\epsinn}{\ensuremath{e^{-j\psi_0^{(N)}}}}
\newcommand{\epsioo}{\ensuremath{e^{-j\psi_1^{(1)}}}}
\newcommand{\epsiok}{\ensuremath{e^{-j\psi_1^{(k)}}}}
\newcommand{\epsion}{\ensuremath{e^{-j\psi_1^{(N)}}}}
\newcommand{\dG}{\ensuremath{\Delta G}}

\newcommand{\dt}{\ensuremath{\Delta t}}
\newcommand{\df}{\ensuremath{\Delta f}}

\begin{document}
	
	\title{Joint Ranging and Clock Synchronization for Dense Heterogeneous IoT Networks}

	\author{\IEEEauthorblockN{Tarik Kazaz, Mario Coutino, Gerard J. M. Janssen, Geert Leus and Alle-Jan van der Veen \thanks{This research was supported in part by NWO-STW under contract 13970 (``SuperGPS''). Mario Coutino is partially
supported by CONACYT.}}

	\IEEEauthorblockA{Faculty of Electrical Engineering, Mathematics and Computer Science\\
			          Delft University of Technology, 2628 CD Delft, The Netherlands
			        }}
	
	\maketitle
	
	\noindent 
	\begin{abstract}
		Synchronization and ranging in internet of things (IoT) networks are challenging due to the narrowband nature of signals used for communication between IoT nodes. Recently, several estimators for range estimation using phase difference of arrival (PDoA) measurements of narrowband signals have been proposed. However, these estimators are based on data models which do not consider the impact of clock-skew on the range estimation. In this paper, clock-skew and range estimation are studied under a unified framework. We derive a novel and precise data model for PDoA measurements which incorporates the unknown clock-skew effects. We then formulate joint estimation of the clock-skew and range as a two-dimensional (2-D) frequency estimation problem of a single complex sinusoid. Furthermore, we propose: (i) a two-way communication protocol for collecting PDoA measurements and (ii) a weighted least squares (WLS) algorithm for joint estimation of clock-skew and range leveraging the shift invariance property of the measurement data. Finally, through numerical experiments, the performance of the proposed protocol and estimator is compared against the Cram\'er Rao lower bound demonstrating that the proposed estimator is asymptotically efficient. 
	\end{abstract}
	
	\IEEEpeerreviewmaketitle
	
	\section{Introduction}
	\noindent Synchronization and localization are key requirements of future internet of things (IoT) applications. IoT networks enable distributed information processing tasks such as sensing, aggregation, and other tasks which benefit from node location information and network-wide synchronization \cite{wu2011clock, win2018efficient}. Typically, IoT nodes are low-power devices equipped with low-cost reference clock sources, i.e. local oscillators, and narrowband radio chips. The individual clocks of the nodes drift from each other due to local oscillator imperfections, environmental and voltage variations. Therefore, it is essential to periodically synchronize and calibrate the clocks in order to keep the time synchronization among the nodes in the network. 
	Clock drifts have a direct impact on various IoT applications, for instance, on the range estimation between the nodes, which is a crucial input for most network localization techniques.
	
	Clock synchronization and node localization in IoT networks have received considerable attention in the past. Many research efforts have approached these problems as either separate or joint estimation problems \cite{wang2011robust, rajan2011joint, chepuri2013joint, rajan2015joint, pelka2014accurate, von2016no, oshiga2016superresolution}. Existing methods can be classified into (i) time-stamping methods based on ultra-wideband (UWB) signals \cite{wang2011robust, rajan2011joint, chepuri2013joint, rajan2015joint}, and (ii) phase-based methods which utilize carrier phase measurements of narrowband signals \cite{pelka2014accurate, von2016no, oshiga2016superresolution}. Methods falling in the first class offer high timing resolution and a plethora of protocols and algorithms for joint ranging and clock synchronization have been proposed \cite{rajan2011joint, chepuri2013joint, rajan2015joint}. However, in general, these methods are not applicable to IoT networks due to the narrowband radio constraints of the nodes. On the other side, the methods based on phase-based ranging, i.e. phase difference of arrival (PDoA), do not consider the impact of unknown clock-skew on range estimation \cite{pelka2014accurate, von2016no, oshiga2016superresolution}. These methods are based on a simplified and inaccurate data model which results in a biased range estimation due to the influence of clock-skew.
	
	In this paper, we aim for joint clock-skew and range estimation from PDoA measurements. We derive a novel and precise data model which considers hardware imperfections of the IoT nodes, i.e. clock-skew of the local oscillators, and wireless channel effects. Therefore, we propose a two-way communication protocol for collecting PDoA measurements over a two-dimensional (2-D) set of equispaced time epochs and carrier frequencies. With this data, a matrix whose rows collect measurements acquired on the same carrier frequencies but different time epochs is constructed. This data matrix exhibits a structure that allows 2-D frequency estimation techniques. Furthermore, we show that the data matrix is rank one and that its principal singular vectors have a shift invariance property which enables joint estimation of the clock-skew and range.
	
	We propose an algorithm for joint clock-skew and range estimation based on weighted least squares (WLS) using the ideas of 2-D frequency estimation \cite{rouquette2001estimation, so2006generalized, so2010efficient}. In this approach, the shift invariance of the left singular vector provides the range estimate, while the shift invariance of the right singular vector provides the clock-skew estimate.
	Finally, the performance of the proposed protocol and estimator is compared against the Cram\'er Rao lower bound (CRLB) using numerical simulations demonstrating that the proposed estimator is asymptotically efficient and approaches the CRLB for sufficiently high signal-to-noise ratio (SNR).
	
	\noindent \textit{Notation:}
	Upper (lower) bold face letters are used to denote matrices (column vectors), while $(.)^T$, $(.)^H$, $(.)^*$, $\odot$, $\bI_{N}$ and $\bzero_{N}$ respectively represent transpose, Hermitian transpose, complex conjugate, element-wise Hadamard product, $N \times N$ identity matrix and $N \times 1$ vector of zeros. Furthermore, $\widehat{(.)}$ denotes estimate of a parameter, $\bbE(.)$ is the expectation operator and $\text{vec(.)}$ forms a vector from a matrix by stacking the columns of the matrix.
	 
	\section{Problem Formulation and System Model}
\begin{figure}[t!]
	    \setlength\belowcaptionskip{-1.2\baselineskip}
		\centering
		\includegraphics[scale=0.55]{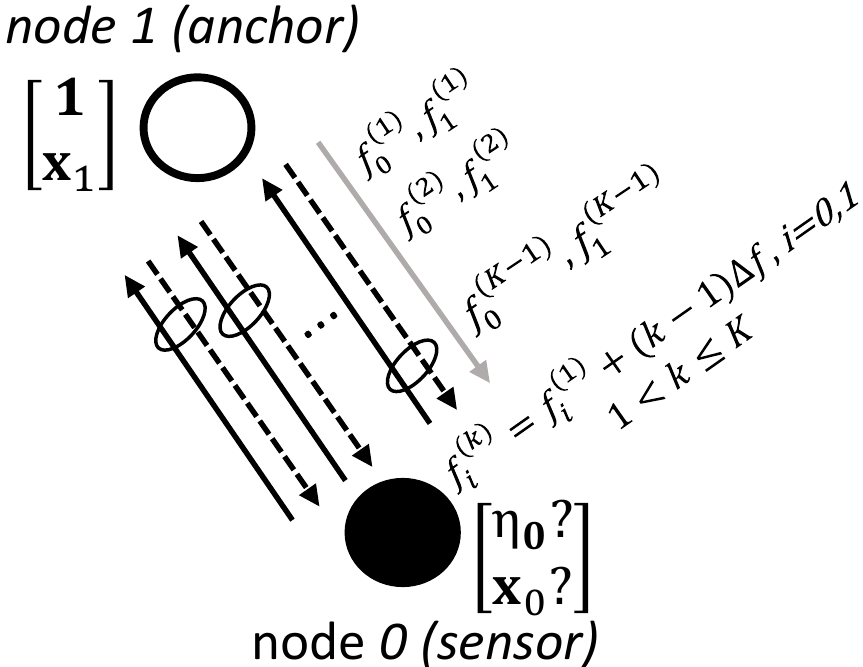}
		\caption{Illustration of the nodes in the IoT network, with known and unknown parameters, two-way carrier messages and a data message.} 
		\label{fig:netmodel}
	\end{figure}

	\noindent Without loss of generality, consider a single sensor (\textit{node 0}) and anchor node (\textit{node 1}) in a fully asynchronous wireless IoT network, as shown in Fig. \ref{fig:netmodel}. Let us, assume that the anchor node has a relatively stable clock oscillator and known position, while the sensor node has an unknown position and a non-ideal oscillator with frequency drift. The clock behavior of the sensor node is considered to be characterized by the first-order affine clock model \cite{tzoreff2014single}
	\begin{equation}
		\label{eq:af_model}
		\nu_0 = \nu_1(1+\eta_o),
	\end{equation} 
	where $\eta_o$ is the clock-skew of the sensor node measured in parts per million (ppm), while $\nu_0$ and $\nu_1$ are the frequencies of the oscillator signals at the sensor and anchor node, respectively.
	
	Here, we assume that the nodes are equipped with narrowband radio transceivers allowing two-way communication. In addition, the radio transceivers support estimation of the phase difference between the carrier frequency of the received signal and its own local oscillator frequency.
	
	For simplicity, consider that the nodes are distributed over a two-dimensional space. Let the vectors $\bx_{i} \in \bbR^{2 \times 1}$, $i = 0, 1$ collect the coordinates of the nodes, where the coordinates of the sensor node $\bx_{0}$ are unknown. The range between the anchor and sensor node is defined as ${d_{01} = d_{10} = \Vert \bx_{1} - \bx_{0} \Vert_{2}}$, where $\Vert \cdot \Vert_{2}$ denotes the Euclidean norm.
	
	
	\noindent \textbf{Frequency synthesizer model.} In order to transmit a signal in a desired frequency band, each radio transceiver generates a carrier signal. The carrier signal is generated by the frequency synthesizer of the transceiver which is driven by the clock signal of the local oscillator. Modern radio transceivers support communication on a number of carrier frequencies which can be selected by changing the gain of the divider in the frequency synthesizer. 
	
	Therefore, we can assume that all the nodes have the same frequency synthesizer with a set of $K$ equispaced gains defined as $\Gk = \Go + (k-1)\dG$, $k= 1, \ldots, K$,  
	where $\Gk \in \bbQ$ is the $k$th gain and $\dG$ is the step of the frequency divider. The carrier frequency generated at the output of the frequency synthesizer for the $k$th gain is given by $\fik=\Gk \nu_i$, $i = 0, 1$ \cite{Razavi:2011:RM:2132691}. The set of all equispaced carrier frequencies supported by the frequency synthesizer can be written as
	\begin{equation}
		\label{eq:freq_set}
		\cF_i = {\lbrace \fik= \fio + (k-1) \df_i : {\lbrace\fio, \df_i \rbrace} \in \bbR \rbrace}_{k=1}^{K},
	\end{equation}
	\noindent where $\df_i = \dG \nu_i$, $i = 0, 1$ is the step of the frequency synthesizer, and it depends on the clock oscillator signal frequency (\ref{eq:af_model}).
	
	\noindent \textbf{Signal model.} Consider that the sensor node transmits a single tone unmodulated carrier signal at the $k$th carrier frequency
	\begin{equation}
		\label{eq:tx_signal}
		\snk(t) = \Re \left \{ \us_0 e^{j\left(2\pi \fnk t+ \phnk\right)} \right \},
	\end{equation}
	\noindent where $\us_0 \in \bbR$ is the amplitude of the complex envelope of $\snk(t)$ and $\vph_0^{(k)}$ is the unknown phase offset introduced by the process of switching the carrier frequency \cite{tse2005fundamentals}. 
	
	The transmitted signal $\snk(t)$ is narrowband, and therefore, it is reasonable to consider flat-fading effects in the channel model. The signal received at the anchor node after propagation through the channel and down-conversion by $\fok$ is given by 
	\begin{equation}
		\label{eq:rx_signal}
		\rnok(t) =  \beta_{01}^{(k)} \us e^{j \left( 2\pi \munok t + \deltanok \right) } + n_1^{(k)}(t),
	\end{equation}
	\noindent where $\beta_{01}^{(k)} \in \bbC$ is the complex path attenuation of the channel at $\fnk$, $\munok = \fnk-\fok$ and $\deltanok = \phnk-\phok$ are the unknown $k$th carrier frequency and phase offsets, respectively, while $n_1^{(k)}(t) \backsim \mathcal{CN} (0,\sigma_1^{2})$ denotes the zero-mean complex Gaussian noise present at the anchor node. The complex path attenuation is defined as $\beta_{01}^{(k)}  = \alpha_{01}^{(k)} e^{-j 2\pi \fnk \tau_{01}}$ where $\alpha_{01}^{(k)} \in \mathbb{R}_+$ is the channel attenuation, ${\tau}_{01}=d_{01}/c = d_{10}/c$ is the unknown propagation delay between two nodes and $c$ is the known propagation speed of the radio signal. Using the frequency synthesizer and clock models, the carrier frequency offset is given by ${\munok = \munoo + (k-1)\Delta \mu_{01}}$ where ${\Delta \mu_{01} = \df_1\eta_o}$.
	
	The objective in this paper is then to estimate the unknown parameters $\eta_o$ and $d_{01}$ given the two-way communication between nodes and PDoA functionalities of the radio transceivers. 
	\section{Communication Protocol and Data Model}
	\noindent In the following, we first derive a detailed data model for PDoA measurements considering a classical two-way protocol for ranging. Then, based on the derived model we propose a novel 2-D PDoA protocol for joint ranging and synchronization.   
	\subsection{Classical PDoA ranging protocol }
\begin{figure*}[t] \setlength\belowcaptionskip{-1.2\baselineskip} \centering \subfloat[]{%
		\includegraphics[trim=0 0 0
		0,clip,width=5.6cm]{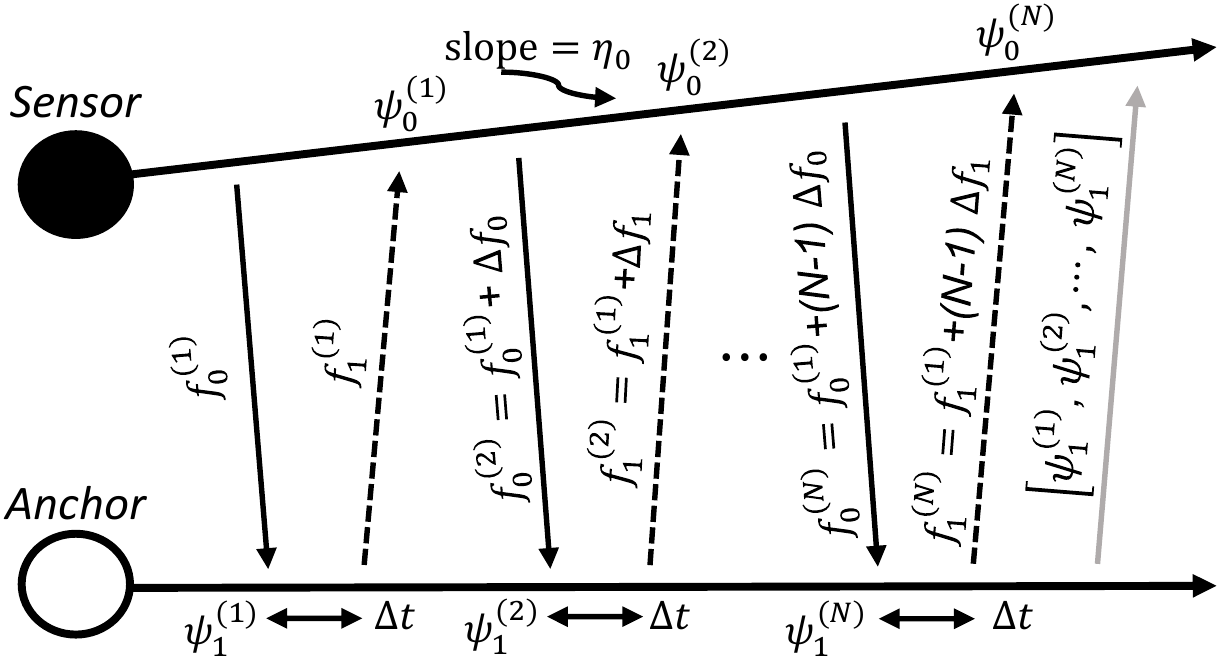}%
		\label{fig:per:a}%
	}\qquad \subfloat[]{%
		\includegraphics[trim=0 0 0
		0,clip,width=7.1cm]{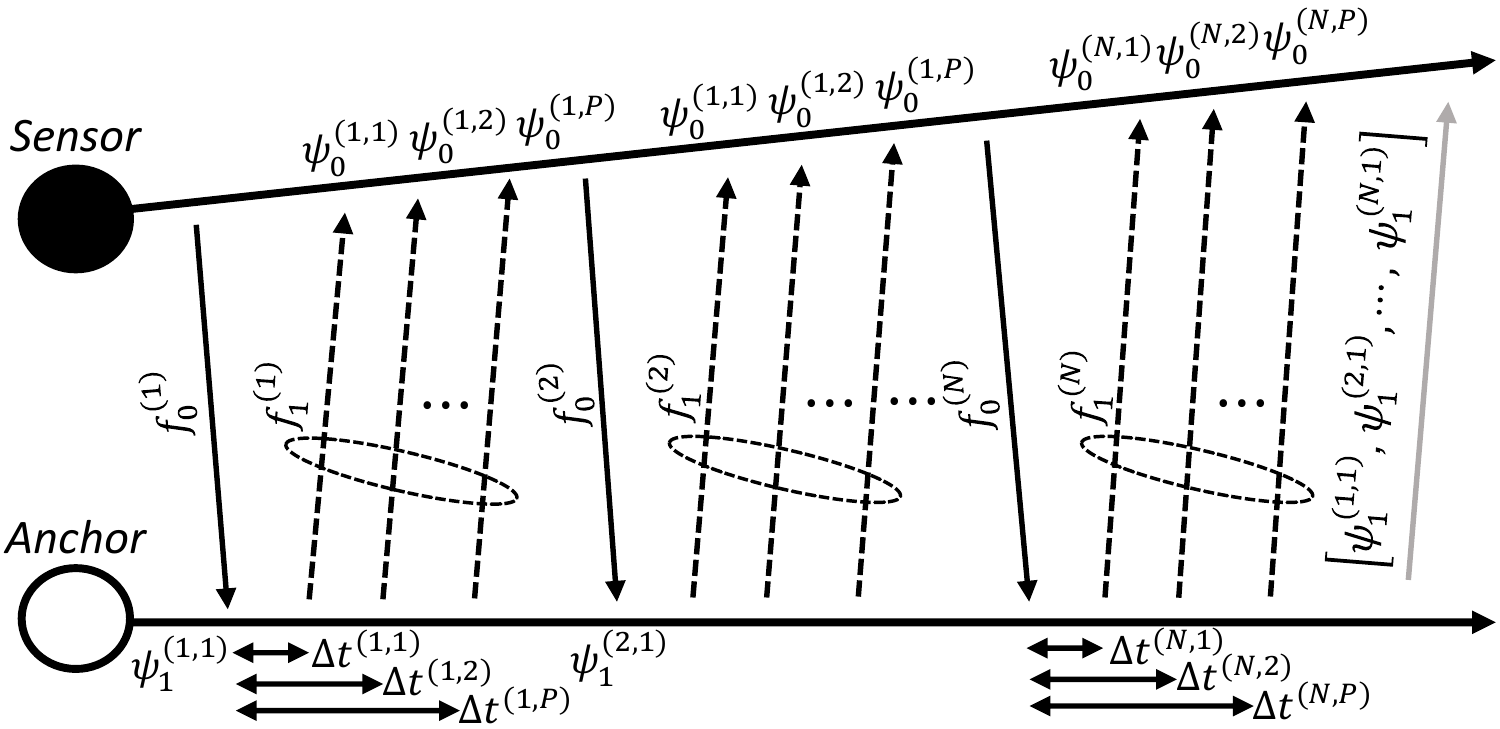}%
		\label{fig:per:b}%
	}\qquad \subfloat[]{%
		\includegraphics[trim=0 0 0
		0,clip,width=4.15cm]{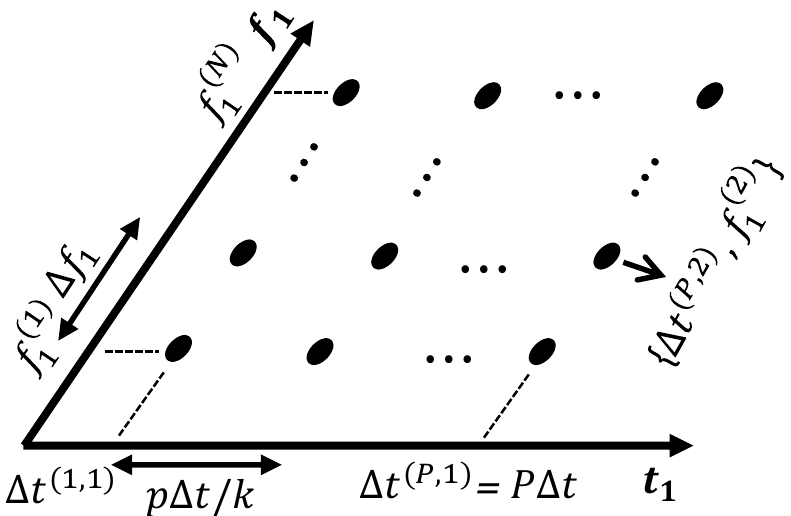}%
		\label{fig:per:c}%
	} \caption{(a) Classical PDoA two-way ranging protocol, (b) 2-D PDoA protocol for ranging and synchronization, and (c) 2-D equispaced time-frequency grid at anchor node.}
	\label{fig:protocols} 
   \end{figure*}

	\noindent In the classical two-way PDoA protocol (cf. Fig. \ref{fig:per:a}) the sensor node initiates the communication and sends a message using the signal $\snk(t)$, i.e. using carrier frequency $\fnk$, to the anchor node. Then, the anchor node receives the message as the signal $\rnok(t)$ and replies back to the sensor node by sending a message using signal $\sok(t)$, i.e. using carrier frequency $\fokm$. After the exchange, both nodes change their carrier frequencies to $\fikp = \fik + \df_i$, $i=0, 1$, and the same two-way message exchange pattern is repeated. The phase difference of the carrier signals, $\psink$ and $\psiok$, using the $k$th carrier frequency are measured at both sensor and anchor nodes, respectively. 
	Now, considering the noiseless case and assuming that channel reciprocity\footnote{The received signals at anchor and sensor nodes differs only in the signs of phase and frequency offset.} conditions hold, using (\ref{eq:rx_signal}) $\psink$ and $\psiok$ are given by
	\begin{equation}
		\label{eq:phase_meas}
		\begin{aligned}
		\psink & =  -2\pi \munok \dt - 2\pi \fok \tau_{01} -\deltanok\\
		& \psiok =  -2\pi \fnk \tau_{01} + \deltanok,
		\end{aligned}
	\end{equation}
	\noindent where $\Delta t$ is the deterministic time epoch between measurements collected at anchor and sensor nodes, while all other nondeterministic timing differences between nodes are absorbed in $\deltanok$. In general, the time epoch $\Delta t$ is controllable by the anchor node and it has values in the order of tens of microseconds. In the classical PDoA protocol, it is assumed that $\Delta t$ is fixed during the recollection of the measurements.
	
	In this paper, we focus on indoor localization scenarios where the channel coherence time is typically of the order of several hundreds of milliseconds \cite{rahul2012jmb}. Hence, we can assume that $N \leq K$ two-way messages have been exchanged according to the PDoA protocol within the channel coherence time. For the sake of simplicity, the $N$ phase difference measurements recorded at sensor and anchor nodes are transformed in their negative complex exponential form and collected in the vectors ${\bb_0 = \left[\epsino, \ldots, \epsinn \right]^T \in \bbC^{N \times 1}}$,  ${\bb_1 = \left[ \epsioo, \ldots, \epsion \right]^T \in \bbC^{N \times 1}}$. 
	
	For ranging purposes, the phase offset represent nuisance parameter which can be eliminated from the acquired measurements by considering ${\ba = \bb_0 \odot \bb_1}$ instead. The argument of the $k$th element in $\ba$ is given by
	\begin{equation}
	\label{eq:sim_diff}
		\text{arg}\{a_k\} = 2\pi \munok \dt + 2\pi (\fok + \fnk) \tau_{01}.
	\end{equation}
	\noindent Using the frequency synthesizer model and ($\ref{eq:phase_meas}$), we can write
	\begin{equation}
		\label{eq:simp_diff1}
	    \begin{aligned} \centering
			&\munok = G^{(1)} \nu_1 \eta_{0} + (k-1) \eta_{0} \df_1, \\
			\fok + & \fnk = (2+\eta_{0}) (G^{(1)} \nu_1 + (k-1) \df_1).
		\end{aligned}
	\end{equation}
	\noindent Therefore, the vector $\ba$ has the model
    \begin{equation}
		\label{eq:pdoa_vector}
		\ba(\tau_{\eta}) = a(\tau_{\eta})\left[1, e^{j2\pi \df_1 \tau_{\eta}},\ldots,e^{j2\pi (N-1) \df_1 \tau_{\eta}}\right]^T,
	\end{equation}
	\noindent where ${a(\tau_{\eta}) = e^{j2\pi G^{(1)}\nu_1\tau_{\eta}}}$ is the the first element in $\ba(\tau_{\eta})$ and ${\tau_{\eta} = \eta_o \dt + (2+\eta_o)\tau_{01}}$. Note that $\ba(\tau_{\eta})$ has a shift invariance structure. This structure is precisely the one that has a uniform linear array (ULA) response vector in array processing \cite{krim1996two}. However, in this case the phase shift of the elements in $\ba(\tau_{\eta})$ is caused by equispaced carrier frequency switching, i.e. frequency hopping. 
	
	\subsection{2-D PDoA ranging and synchronization protocol}
	\noindent The shift invariance of $\ba(\tau_{\eta})$ only allows for the estimation of a single parameter $\tau_{\eta}$. However,  $\eta_o$ and $\tau_{01}$, i.e. $d_{01}$, cannot be uniquely determined from $\ba(\tau_{\eta})$. For example, estimation of the $\tau_{01}$ from $\ba(\tau_{\eta})$ results in an estimate biased by the clock-skew. To alleviate this, here, we are interested in a protocol for collecting measurements that allows joint clock-skew and range estimation. 
	
	In the classical PDoA protocol, measurements are collected over the set of equispaced carrier frequencies while the time epoch $\Delta t$ is fixed during message exchange. In the 2-D PDoA protocol, we propose to collect the measurements over a 2-D set of equispaced time epochs and carrier frequencies (cf. Figs. \ref{fig:per:b} and \ref{fig:per:c}). In this case, the sensor node transmits a single message per two-way exchange, while the anchor node transmits $P$ messages based on the equispaced time epochs. The set of the equispaced time epochs for the $k$th carrier frequency is given by $\Delta t^{(k,p)} = p\Delta t/k$, where $k=1,\ldots,N$ and $p=1,\ldots,P$. Note that these time epochs depend on the index of the carrier frequency, i.e. $k$. 
	
	The $P$ phase difference measurements recorded at the sensor node for the $k$th carrier frequency are transformed in their negative complex exponential form and collected in the vector $\bb_k \in \bbC^{P \times 1}$. As before, we follow a similar approach for nuisance parameters elimination. The vector that collects noiseless PDoA measurements recorded at the $k$th carrier frequency is written as ${\ba_k =\epsiok\bb_k \in \bbC^{P \times 1}}$ which satisfies the model
	\begin{equation}
		\label{eq:2pdoa_vector}
		\ba_k(\eta_o, \tau_{01}) = a(\tau_{\eta})\gamma^{k-1}[1, \phi, \ldots, \phi^{P-1}]^T,
	\end{equation}
	\noindent where $a(\tau_{\eta})$ is defined in (\ref{eq:pdoa_vector}), $\gamma =  e^{j2\pi \df_1(2+\eta_o)\tau_{01}}$ and $\phi = e^{j2 \pi \df_1 \eta_o \dt}$.
		
	\noindent \textbf{Remark:} (Practical implementation): The 2-D PDoA protocol requires that during a single two-way message exchange no carrier frequency switching occurs. This constraint ensures that the phase offset between two nodes remains constant during time hopping. However, there is no constraint on the frequency hopping sequence. This makes the proposed protocol attractive for implementation as an adaption of existing medium access control protocols such as time-slotted channel hopping (TSCH) or WirelessHART \cite{watteyne2015using}.         
	
	
    
    \section{Joint Clock-skew and Range Estimation}	
    \noindent In the following, we show how to jointly estimate clock-skew, i.e. $\eta_o$, and range, i.e. time delay $\tau_{01}$, from collected measurements.
    
    The noise-corrupted version of $\ba_k$ is given by $\bmm_k = \ba_k + \bn_k$, where $\bn_k$ is a zero-mean complex Gaussian distributed noise vector\footnote{The phase estimation errors in the PLLs are Thikonov, i.e. von Mises distributed \cite{shmaliy2005mises}. However, for large signal to noise ratio, the Thikonov distribution can be approximated by a Gaussian distribution.}. From a set of $N$ noisy 2-D PDoA measurements, we construct a measurement matrix of size $P \times N$ as
    \begin{equation}
    	\label{eq:mes_matrix}
     	\bM = [\bmm_1, \ldots, \bmm_N].
    \end{equation}
    The measurement matrix satisfies the model
    \begin{equation}
    	\label{eq:mes_matrix2}
    	\bM = \bA + \bN,
    \end{equation}
    where $\bA = [\ba_1, \ldots, \ba_N]$ and ${\bN \in \bbC^{P \times N}}$ is the noise matrix. Using (\ref{eq:2pdoa_vector}), it is straightforward to show that $\bA$ can be modeled as
    \begin{equation}
	    \label{eq:data_model}
	    \bA = \bq(\eta_{0}, \tau_{01})\bh^T(\eta_{0}, \tau_{01}),
	\end{equation}
	\noindent where 
    \begin{equation}
	    \label{eq:mes_model}
	    \begin{aligned}
	    \bq=&a(\tau_{\eta})[1, \phi, \ldots, \phi^{P-1}]^T\\
	    \bh&=[1, \gamma, \ldots, \gamma^{N-1}]^T.
	    \end{aligned}
	\end{equation}
    \noindent Model (\ref{eq:mes_matrix2}), using relation (\ref{eq:data_model}),  resembles the signal model for 2-D frequency estimation of a single complex sinusoid in white Gaussian noise. This is a classical signal processing problem for which numerous methods have been proposed \cite{clark1994two,rouquette2001estimation, so2006generalized, so2010efficient}. Although the maximum likelihood estimator proposed in \cite{clark1994two} can attain optimum performance, it has high computational requirements due to the multidimensional search. Here, we are interested in suboptimal but computationally more attractive (practical) methods. To do so, inspired by~\cite{so2010efficient}, we develop an algorithm for joint clock-skew and range estimation. 
    
    From \eqref{eq:data_model}, we can observe that $\bA$ has rank one and that the vectors $\bq$ and $\bh$ span its column and row space, respectively. Since $\bq$ and $\bh$ exhibit shift invariance, it is possible to estimate $\gamma$ and $\phi$ from the low-rank approximation of $\bM$. Then, from $\phi$ and $\gamma$, the parameters $\eta_o$ and $\tau_{01}$, i.e. $d_{01}$, immediately follow. 
    
    \begin{figure*}[h] \setlength\belowcaptionskip{-1.2\baselineskip} \centering
    	\subfloat[]{%
    		\includegraphics[trim=0 0 0
    		0,clip,width=6.0cm]{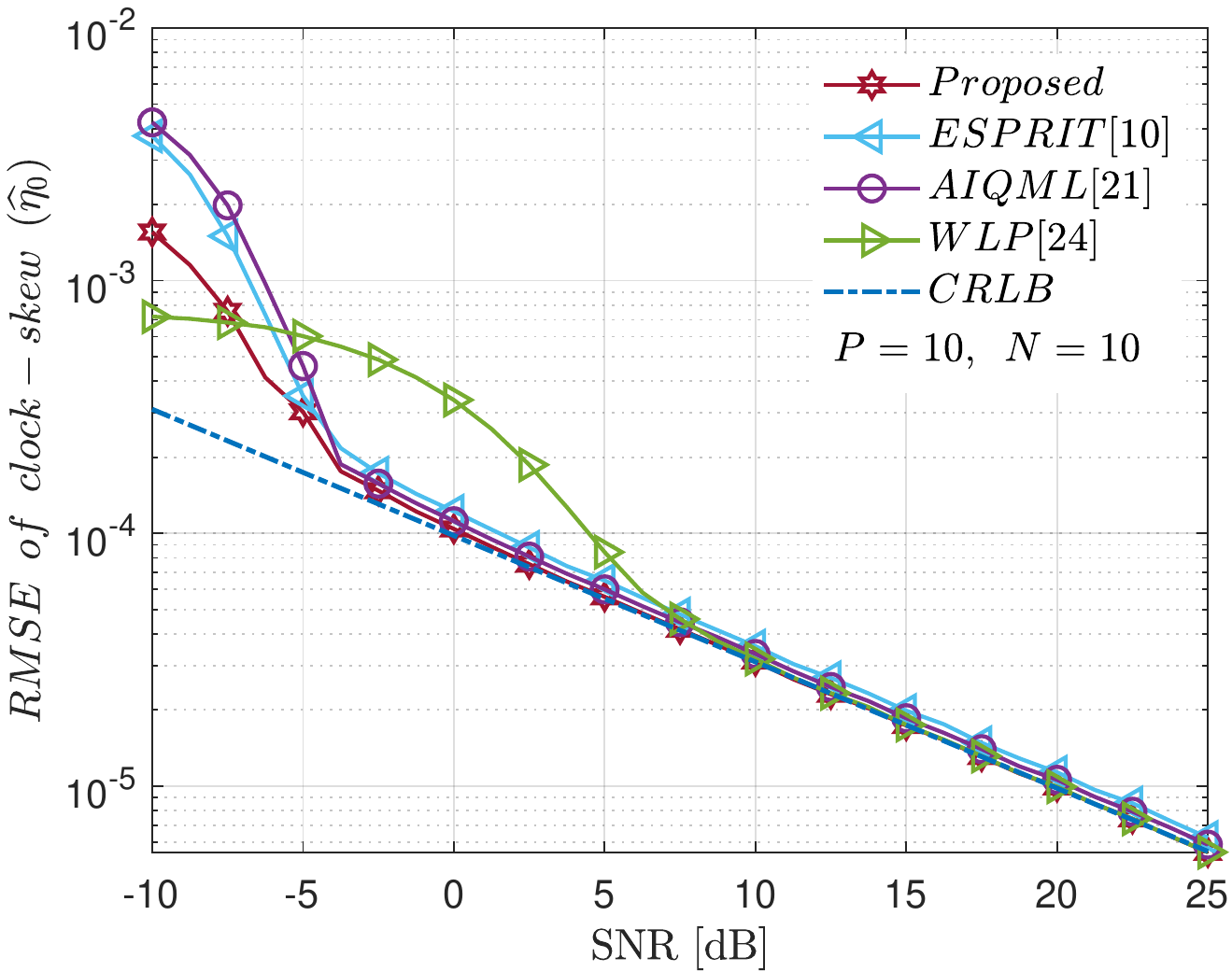}%
    		\label{fig:resw:a}%
    	}\hfil \subfloat[]{%
    		\includegraphics[trim=0 1 0
    		0,clip,width=6.0cm]{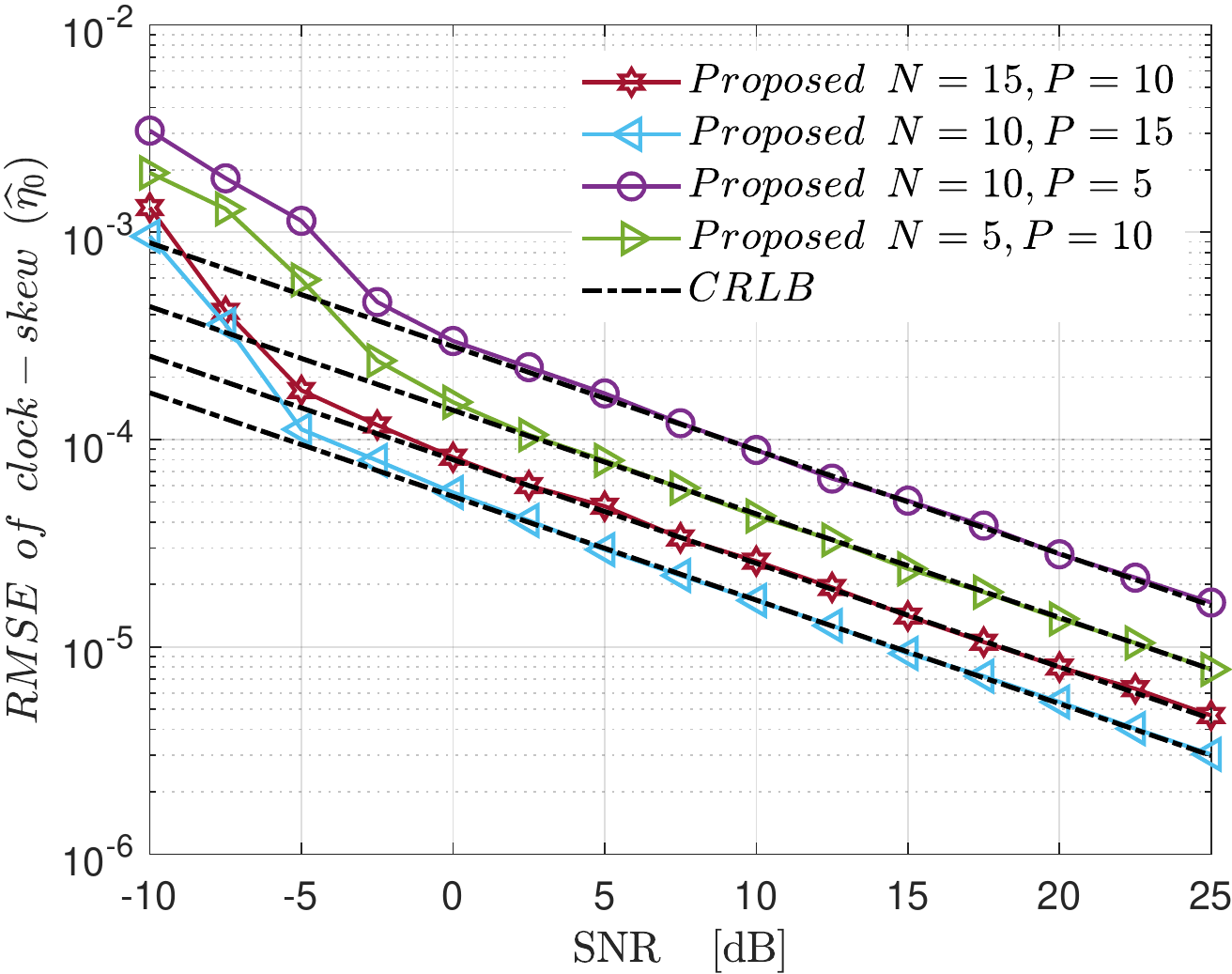}%
    		\label{fig:resw:b}%
    	}\hfil  \subfloat[]{%
    		\includegraphics[trim=0 1 0
    		0,clip,width=6.0cm]{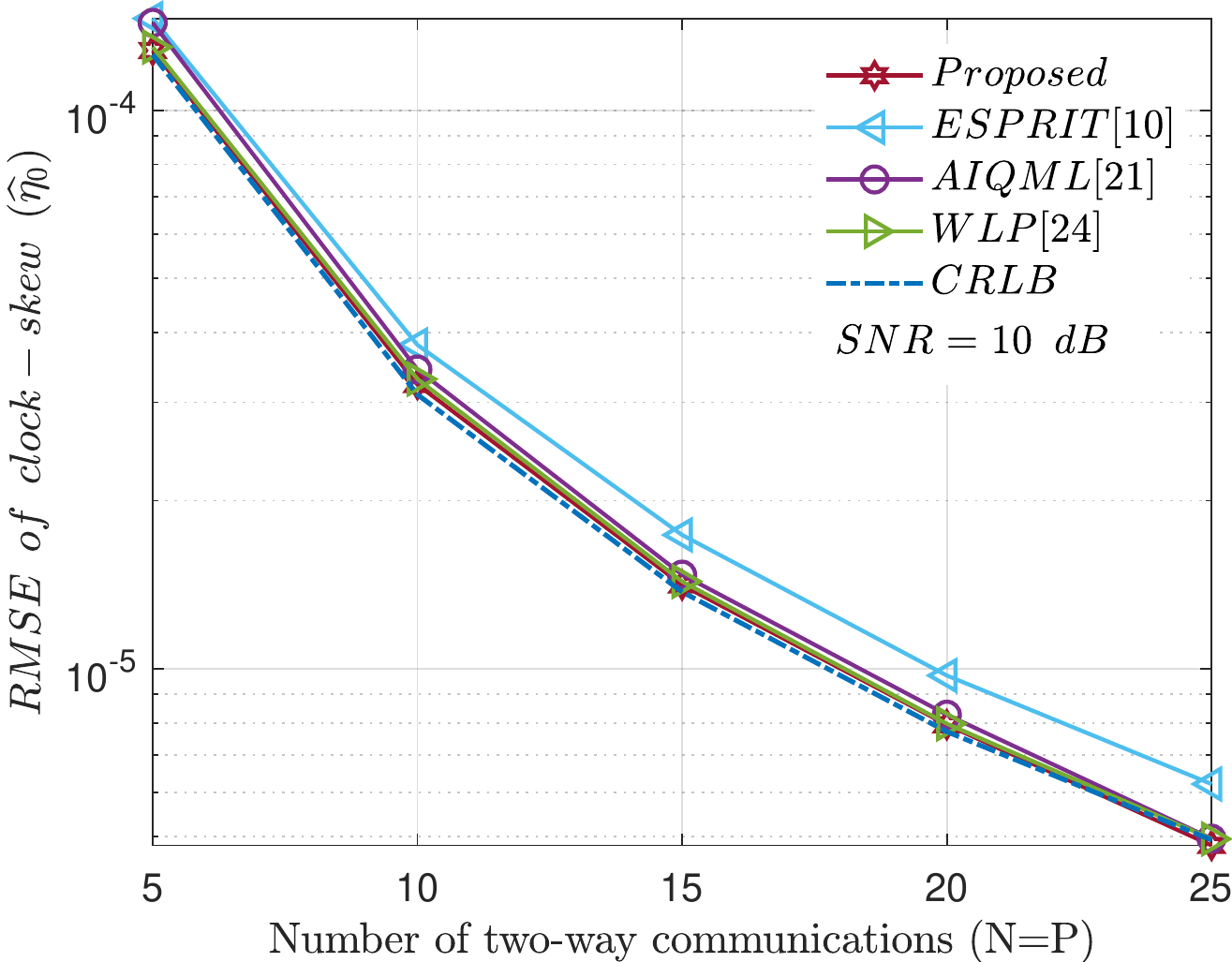}%
    		\label{fig:resw:c}%
    	} \caption{RMSE of estimated clock-skew, $\widehat{\eta}_{0}$, vs (a-b) signal to noise ratio, and (c) number of two-way communications.}
    	
    	\label{fig:resd} 
    \end{figure*}
    \begin{figure*}[h] \setlength\belowcaptionskip{-1.2\baselineskip} \centering \subfloat[]{%
    		\includegraphics[trim=0 0 0
    		0,clip,width=6.0cm]{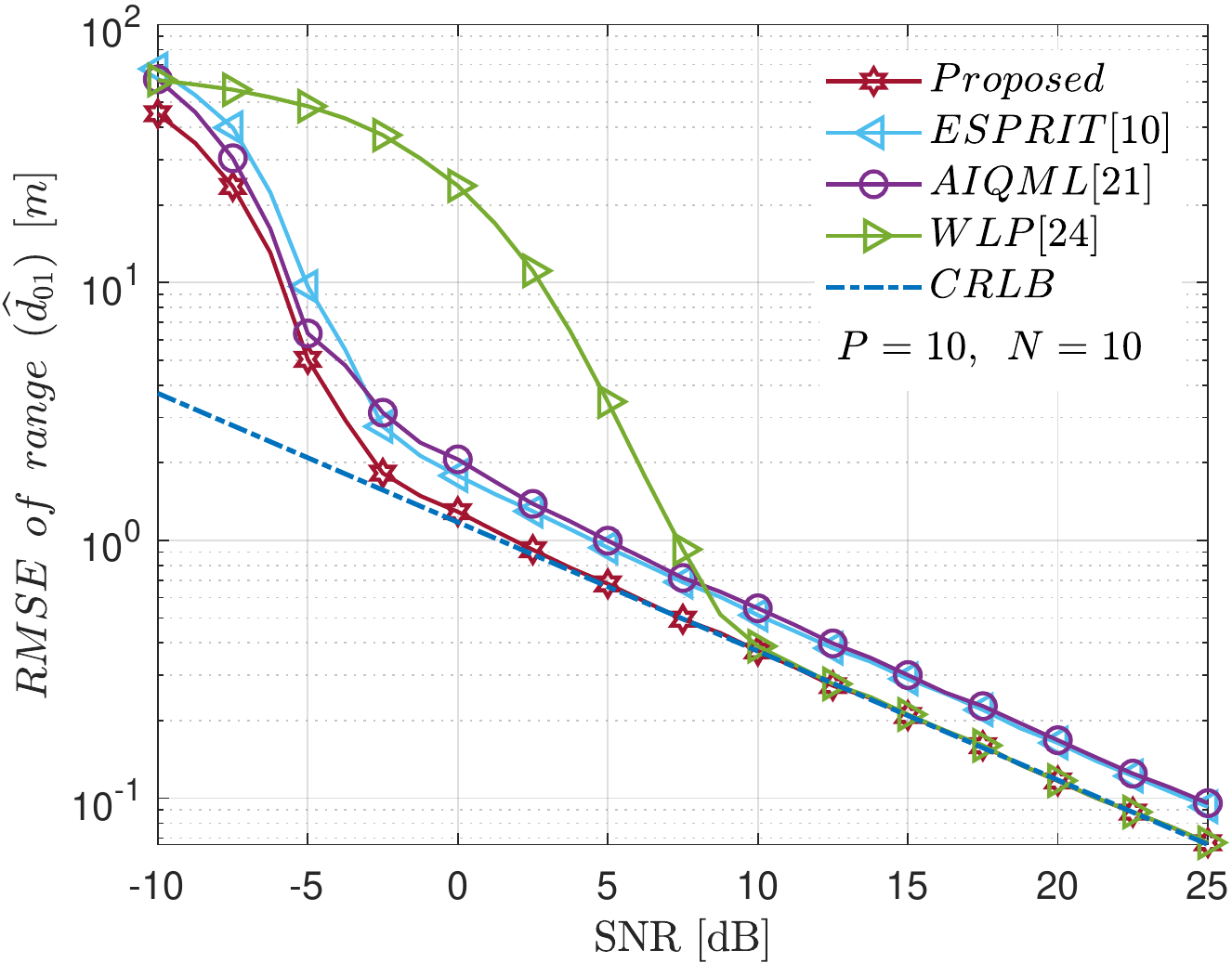}%
    		\label{fig:resd:a}%
    	}\hfil \subfloat[]{%
    		\includegraphics[trim=0 1 0
    		0,clip,width=6.0cm]{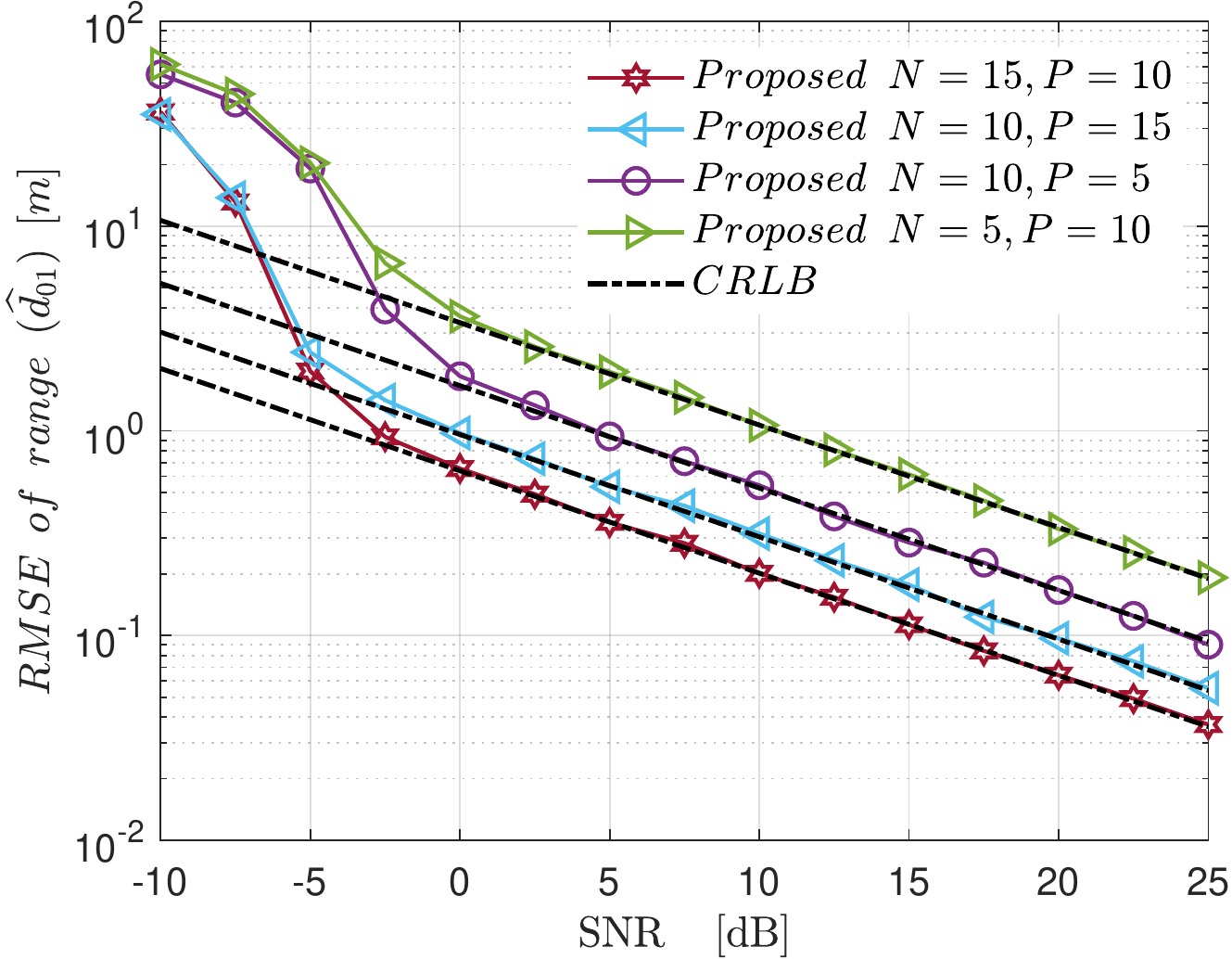}%
    		\label{fig:resd:b}%
    	}\hfil  \subfloat[]{%
    		\includegraphics[trim=0 1 0
    		0,clip,width=6.0cm]{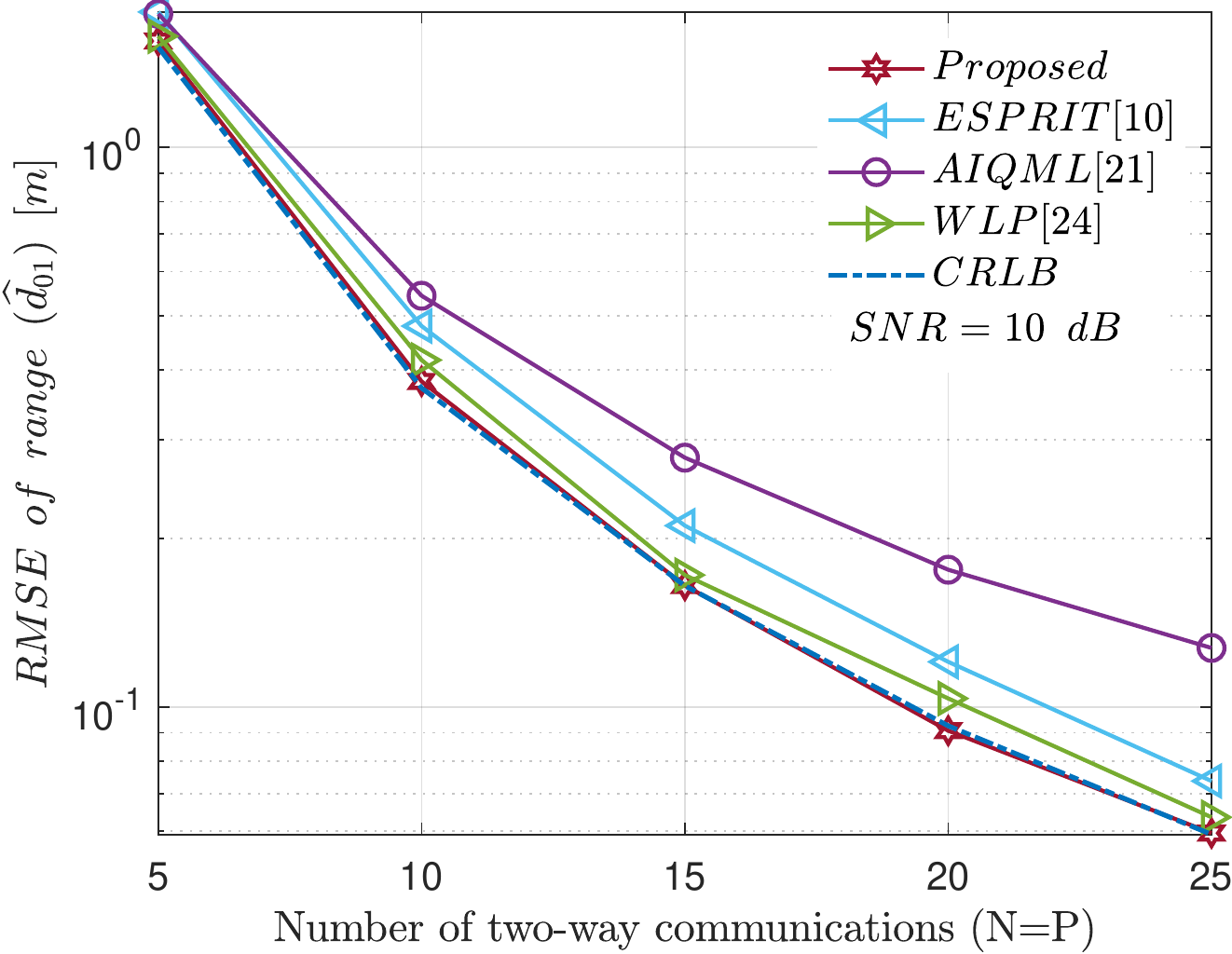}%
    		\label{fig:resd:c}%
    	} \caption{RMSE of estimated range, $\widehat{d}_{01}$, vs (a-b) signal to noise ratio, and (c) number of two-way communications.}
        \label{fig:resd} 
     \end{figure*}
    
    In particular, let $\bu_1$ and $\bv_1$ be the principal orthonormal basis vectors for the column and row span of the rank-one approximation of $\bM$, respectively. These vectors can be obtained using the singular value decomposition (SVD) of $\bM$ and can be expressed as
    \begin{equation}
    \label{eq:vec_rel}
    \bu_1=1/\rho_q \bq, \qquad \bv_1 = 1/\rho_h{\bf h}^*,
    \end{equation}
    \noindent where $\rho_q$ and $\rho_h$ are unknown complex constants. Now, let us define the selection matrices:    
	\begin{equation} 
		\label{eq:sel_mat2}
		\begin{aligned} \centering 
		\bJ_{\phi1} = [\bI_{P-1} \quad \bzero_{P-1}], 
		& \qquad
		\bJ_{\gamma1} = [\bI_{N-1} \quad \bzero_{N-1}], 
		\\
		\bJ_{\phi2} = [\bzero_{P-1} \quad \bI_{P-1}], 
		& \qquad 
		\bJ_{\gamma2} = [\bzero_{N-1} \quad \bI_{N-1}].  
		\end{aligned} 
	\end{equation}
	\noindent To estimate $\phi$, we take subvectors consisting of the first and, respectively, the last $P-1$ elements of the $\bu_1$. That is, we consider ${\bu_{\phi1} = \bJ_{\phi1}\bu_1}$ and ${\bu_{\phi2} = \bJ_{\phi2}\bu_1}$, respectively. We follow the same process for the estimation of $\gamma$, i.e. we take subvectors ${\bv_{\gamma1} = \bJ_{\gamma1}\bv_1}$ and ${\bv_{\gamma2} = \bJ_{\gamma2}\bv_1}$, respectively. From the shift invariance property of $\bu_1$ and $\bv_1$ we have that
	\begin{equation}
		\label{eq:vec_form}
		\bu_{\phi2} \approx \bu_{\phi1} \phi \quad \text{and} \quad \bv_{\gamma2} \approx \bv_{\gamma1} \gamma^*.
	\end{equation}
	\noindent In the case of white noise, the approximate solutions to the relations in ($\ref{eq:vec_form}$) can be found using least squares (LS). However, here we are adopt the weighted least squares (WLS) approach \cite{so2006generalized} and formulate problem ($\ref{eq:vec_form}$) as
   \begin{equation}
	   \label{eq:wls_prob}
	   \begin{aligned} \centering
	   &\widehat{\phi}= \argminA_{\phi} \Vert \bC_{\phi}^{-1/2}(\bu_{\phi1} \phi - \bu_{\phi2}) \Vert_{2}^2\\
	   &\widehat{\gamma}= \argminA_{\gamma}  \Vert \bC_{\gamma}^{-1/2}(\bv_{\gamma1} \gamma^{*} - \bv_{\gamma2}) \Vert_{2}^2,
	   \end{aligned}
   \end{equation} 
   \noindent where ${\bC_{\phi} =\bbE(\br_{\phi}\br_{\phi}^H)}$ and ${\bC_{\gamma} = \bbE(\br_{\gamma}\br_{\gamma}^H)}$ are the covariance matrices of the residuals ${\br_{\phi}=\bu_{\phi1} \phi - \bu_{\phi2}}$ and ${\br_{\gamma}=\bv_{\gamma1} \gamma^{*} - \bv_{\gamma2}}$, respectively. Therefore, the weighting matrices are the inverse of the covariance of the residuals, i.e. ${\bW_{\phi}=\bC_{\phi}^{-1}}$ and ${\bW_{\gamma}=\bC_{\gamma}^{-1}}$. The optimal $\bW_{\phi}$ and $\bW_{\gamma}$ for the considered problem are given in closed-form by \cite{so2006approximate}
   \begin{equation}
   \label{eq:wls_prob}
   \begin{aligned} \centering
	   &\bW_{\phi}[p,n]= (P\text{min}(p,n)-pn)\phi^{(p-n)}/P\\
	   &\bW_{\gamma}[p,n]= (N\text{min}(p,n)-pn)\gamma^{(n-p)}/N,
   \end{aligned}
   \end{equation}
   \noindent where $p=1,\ldots,P$ and $n=1,\ldots,N$. Note that $\bW_{\phi}$ and $\bW_{\gamma}$ depend on the unknown parameters $\phi$ and $\gamma$. Therefore, first we estimate $\phi$ and $\gamma$ using LS and then these estimates are used for construction of $\bW_{\phi}$ and $\bW_{\gamma}$. Finally, the WLS is used to obtain $\hat\phi$ and $\hat\gamma$. Based on the WLS estimates of $\phi$ and $\gamma$ the unknown parameters are computed as
   \begin{equation}
   \label{eq:wls_prob}
   \begin{aligned} \centering
	   &\widehat\eta_o = (2 \pi \df_1 \dt)^{-1}\text{arg}(\hat\phi)\\
	   \widehat d_{01} &= c(2 \pi \df_1 \dt (2+\hat\eta_o))^{-1}\text{arg}(\hat\gamma).
   \end{aligned}
   \end{equation}   
   \noindent Note that first the clock-skew is estimated and later this estimate is used for the estimate of the range.

    \section{Results}
    \subsection{Cram\'er Rao Lower Bound}
    \noindent To assess the performance of the proposed estimators (\ref{eq:wls_prob}) we derive the CRLB for joint clock-skew and range estimation using the model (\ref{eq:mes_matrix2}). For an unbiased estimator $\widehat{\btheta}$, the CRLB is the lower bound on the error variance, that is 
	\begin{equation}
		\label{eq:crlb1}
		\text{var}(\widehat{\btheta}) \geq \bF^{-1}
	\end{equation}
	\noindent where ${\text{var}(\widehat{\btheta}) = \bbE((\widehat{\btheta}- \btheta)(\widehat{\btheta}- \btheta)^T)}$ and $\bF$ is the Fisher information matrix. We assume that the proposed estimators (\ref{eq:wls_prob}) are approximately unbiased for sufficiently large SNR and well designed measurement matrix $\bM$ \cite{liu2008first}. 
	
	In the case of 2-D frequency estimation of the sum of the sinusoids, the Fisher information matrix is given by \cite{hua1992estimating}
    \begin{equation}
		\label{eq:fisher}
		\bF_{p,k} = 2\bsig^{-2}\Re\left[ \frac{\partial \ba^H}{\partial \theta_p} \frac{\partial \ba}{\partial \theta_k} \right] \in \bbR^{2 \times 2}
	\end{equation}
	\noindent where $\bF_{p,k}$ is the ($p$, $k$)th element of $\bF$, $\sigma^{2}$ is the variance of the noise, $\partial/\partial \theta_p$ is the partial derivative with respect to the $p$th element of $\btheta$, $\ba = \text{vec}(\bA) \in \bbC^{PN \times 1}$ is the vector formed by stacking the columns of $\bA$. The resulting Fisher information matrix is invertible, so closed-form expressions for the CRLBs are given by
	\begin{equation}
		 \label{eq:crlb2}
		 \begin{aligned} \centering
		 &\text{var}(\widehat \eta_{0}) \geq \dfrac{6}{SNR(2\pi \df_1 \dt)^2  PN(P^2-1)}, \\
		 &\text{var}(\widehat d_{01}) \geq \dfrac{6c^2}{SNR(4\pi \df_1)^2  PN(N^2-1)}.
		 \end{aligned}
	 \end{equation} 
	 \noindent where $SNR  = \sigma^{-2}$.

    \subsection{Simulations} 
    \noindent In the following, simulations are used to compare the performance of the proposed protocol and algorithm with state-of-the-art estimators for the same problem. We consider two nodes, i.e. anchor and sensor, which are deployed randomly within a range of $140 m$. The carrier frequency step $\Delta f_i$, $i=0,1$ and time epoch $\Delta t$ are set to $0.5 MHz$ and $80 \mu s$, respectively. The clock-skew of the sensor node $\eta_o$ is set to $80 ppm$. The phase difference of arrival measurements are corrupted with zero-mean Gaussian noise and all results presented are averaged over $10^3$ independent Monte Carlo runs.

    Figs. \ref{fig:resw:a} and \ref{fig:resd:a} show the root mean square error (RMSE) of clock-skew and range against the signal-to-noise-ratio for various estimators. The number of time and frequency hops is equal and set to $10$. All the algorithms are independently applied to the same set of PDoA measurements. As shown in the figures, the proposed algorithm outperforms the approximate iterative quadratic maximum-likelihood (AIQML) \cite{so2006approximate}, weighted linear predictor (WLP) \cite{60114} and ESPRIT \cite{rouquette2001estimation} for both clock-skew and range. Furthermore, for sufficiently high SNR the proposed algorithm is asymptotically efficient and approaches the theoretical bounds, i.e. the CRLB.
    
    Fig. \ref{fig:resw:b} and \ref{fig:resd:b} show the RMSE of clock-skew and range against the SNR for a different number of PDoA measurements collected over time epochs ($P$) and carrier frequencies, ($N$). It is shown that by increasing the number of PDoA measurements collected over time epochs the accuracy of the clock-skew estimates is increased, while the accuracy of the range estimates increases with $N$.
   
    Fig. \ref{fig:resw:c} and \ref{fig:resd:c} show the RMSE of clock-skew and range against the number of PDoA measurements, while SNR is set to $10dB$. In all scenarios the number of time and frequency hops is equal. Similar as in the previous scenarios, the proposed algorithm outperforms AIQML, WLP and ESPRIT. In addition, it can be seen that the proposed estimator achieves the CRLB.  
    
    \section{Conclusions}
    \noindent In this paper, we investigated the problem of joint ranging and clock-skew estimation using PDoA measurements. A novel and precise data model for PDoA measurements is derived. The derived model enables joint clock-skew and range estimation from PDoA measurements collected over a 2-D equispaced time-frequency grid. We have proposed a novel protocol for collection of PDoA measurements and an algorithm based on WLS to jointly estimate the clock-skew and range. The presented algorithm leverages shift invariance properties of principal singular vectors of the collected measurements. The proposed estimator is asymptotically efficient and reaches the CRLB for sufficiently high SNR. 

\bibliographystyle{IEEEtran}

\bibliography{bare_conf}

\end{document}